# Vibrational Spectra of $Pb_2Bi_2Te_3$, $PbBi_2Te_4$ and $PbBi_4Te_7$ Topological Insulators: Temperature Dependent Raman and Theoretical Insight from DFT Simulations

Priyanath Mal[a], Ganesh Bera[a], G. R.Turpu[a], Sunil. K. Srivastava[b], A. Gangan[c], Brahmananda Chakraborty*[c], Bipul Das[d] and Pradip Das[†a]

[a]Department of Pure and Applied Physics, Guru Ghasidas Vishwavidyalaya, Koni, Bilaspur-495009, India.

[b]Department of Physics, Mahatma Gandhi Central University, East Champaran, Motihari, Bihar-845401, India.

[c]High Pressure and Synchroton Radiation Physics Division, Bhabha Atomic Research Center, Trombay, Mumbai-40008, India.

[d]National Changhua University of Education, Jin-De Road, Changhua 500, Taiwan.

## Abstract

We present temperature dependent frequency shift and line broadening of phonon modes by insertion of atomic layers of Pb and PbTe in the prototype 3D topological insulator $Bi_2Te_3$, using Raman spectroscopy. Good quality single crystals of $Pb_2Bi_2Te_3$, $PbBi_2Te_4$ and $PbBi_4Te_7$ are grown using the modified Bridgman technique. The Raman modes show progressive blue shift with the decrease in temperature from 298 K to 93 K in $Pb_2Bi_2Te_3$, $PbBi_2Te_4$ and $PbBi_4Te_7$ is due to anharmonic vibrations of the lattice as well as increasing strength of Bi-Te covalent interactions. Experimental results are complemented by extensive density functional theory calculations where a reasonable matching between experimental and computational data is found. Chemical pressure, induces by the insertion of Pb and PbTe layers in $Bi_2Te_3$, modifies the interactions at the boundaries of the quintuple-layers which is evident from the evolution of $A_{1u}^2$ mode. The enhancement of out-of-plane Bi-Te vibrations with respect to in-plane Bi-Te vibrations are observed at low temperatures.

# Introduction

Topological insulators (TIs) are materials that behave as insulators in its interior, but always have conducting boundaries. This compelling characteristic of topological insulator is possible due to strong spin orbit coupling (SOC) of heavy elements, which constitute the topological insulators. [1, 2, 3] Furthermore, the topological order is protected by time reversal symmetry analogous to the quantum spin Hall effect.[1, 2, 3, 4] The potential applications of TIs in spintronic [5], quantum computing and low power electronics [1, 2] etc. bring it into the spot light of research over the last decade. Considerable amount of research efforts have been devoted to understand the layered structures, e.g. $Bi_2Te_3$, $PbBi_2Te_4$ and $PbBi_4Te_7$ due to their unique and interesting bulk thermo-electric properties [6, 7, 8, 9] and topological surface states. [10, 11, 12, 13] It is crucial to understand the phonon-phonon and electron-phonon interactions in these layered materials, in order to use these fascinating materials for practical use. Recently, phonon confinement effect is identified through Raman scattering below three quintuple-layers (QLs) for $Bi_2Te_3$ by Wang *et al.* [14] Most studied $Bi_2Se_3$, categorically in the $B(V)_2A(VI)_3$ group possesses $D_{3d}^5$ point group symmetry having fifteen different bulk phonon modes at $\Gamma$ point and the irreducible representations can be expressed as: $\Gamma_{bulk} = 2(A_{1g}+E_g)+3(E_u+A_{2u})$.[15] The odd and even parity phonons are related to Raman and IR active modes of lattice vibrations respectively, of the bulk $Bi_2Te_3$ like materials with inversion symmetry in the crystal structure.[16] Out of fifteen phonon modes, three ($E_u$) are acoustic modes and another twelve are optical in which, $2(E_g + A_{1g})$ are Raman active in the range 20-200 cm$^{-1}$.[15] The Raman active $E_g$ modes are related to in-plane A(VI)-B(V) and that of $A_{1u}$ corresponds to out-of-plane IR lattice vibrations.[17] Cheng *et al.* [18] predict $A_{1g}$ and $A_{1u}$ modes are affected by the van-der-Waals interactions present in between QLs of layered $Bi_2Te_3$ and independent of spin orbit coupling. The modes $E_g$ and $E_u$ both have nearly the same frequency split due to the van-der-Waals interactions.[18] Although the room temperature (RT) and low temperature Raman characterizations of the prototype $Bi_2Se_3$ and $Bi_2Te_3$ TIs are available in literature, but little attention has been paid to Pb based TIs such as $PbBi_2Te_4$ and $PbBi_4Te_7$ at room temperature as well as at low temperatures. Topological phenomenon has also been reported in $Bi_2/Bi_2Te_3$ heterostructure.[19] $Bi_4Te_3$ has same structure as that of hetero structure is one of the member among seven different room temperature phases of $Bi_xTe_y$.[20, 21] Lind et al.[22] have theoretically predicted that the insertion of $Bi_2$ layers periodically in the van-der-Waals gaps of $Bi_2Se_3$ closes the bulk energy gap of $Bi_2Se_3$ converts it into a semimetal. Valla et al.[23] have explored the existence of topologically protected surface states at the Γ point in the single crystal samples of $Bi_4Se_{2.6}S_{0.4}$ and found $Bi_4Se_{2.6}S_{0.4}$ as topological semimetal. Many issues are related [20, 24] to the vibrational properties of $(Bi_2)(Bi_2Te_3)$ have not yet received desire amount of attention, motivate us to investigate the vibrational properties of intermetallic $(Pb_2)(Bi_2Te_3)$ single crystal for better understanding of these isostructural compounds, where the topological behavior by angle resolve photoemission spectroscopy (ARPES) is yet to set.

Here we provide a systematic investigation of the synthesis and temperature dependent Raman studies of $Bi_2Te_3$, $Pb_2Bi_2Te_3$, $PbBi_2Te_4$ and $PbBi_4Te_7$ high quality single crystals in the temperature range of ~93 K to 298 K. The observed blue shift in the Raman modes is explained in terms of anharmonic lattice vibrations of $Bi_2Te_3$ and $Pb_2Bi_2Te_3$ and or by increasing stiffness of Bi-Te interactions for $PbBi_2Te_4$ and $PbBi_4Te_7$ at lower temperature. The chemical pressure induces by intercalation results in observation of out-of-plane Bi-Te vibrations at ~132 cm$^{-1}$ for $Pb_2Bi_2Te_3$ and $PbBi_2Te_4$. The observed modes of all the samples are theoretically verified through density functional theory (DFT) calculations.

## Experimental

High quality single crystals of binary $Bi_2Te_3$, ternary $Pb_2Bi_2Te_3$, $PbBi_2Te_4$ and $PbBi_4Te_7$ layered chalcogenides are synthesized by reacting homogeneously mixed powder of Bi (Alfa Aesar), Pb (Sigma-Aldrich) and Te (Sigma-Aldrich) of analytical grade in a sealed evacuated quartz tube with conical tip, flashing frequently with Ar. The ampoules are then heat treated to ~1223 K for 24 hrs. and are cooled it down to ~893 K in an ~6 days followed by furnace cooling to room temperature. During the growth, to avoid multi-nucleation the ampoules are kept in pointed bottom configuration, resulting in the growth of the ingots of suitable length with different large crystalline faces. The crystallinity of the samples are investigated at RT by using Rigaku x-ray diffractometer, operating at 3 kW, equipped with Cu-Kα radiation in the 2θ angular range of 10° to 90° with a step size of 0.01°. The chemical homogeneity of the prepared single crystals is then confirmed through SEM-EDAX analysis. Temperature dependent Raman spectra are then recorded in the basal plane of a back scattering geometry for the same crystalline faces of different specimens in the range of 298 K to ~93 K with Technos STR-500 micro-Raman spectrometer equipped with 532 nm diode laser source with a spectral resolution of 1 cm$^{-1}$. The sample stage from Linkam THMS600 facilitated with temperature controller is used to control the sample temperature at desired value in each step of the temperature dependent Raman (TDR) measurement. During the Raman measurement the laser power is kept < 2 mW in order to avoid surface burning and oxidation.

## Results and discussion

As grown silver color platelet crystals are easily cleaved off from the ingots. For polycrystalline x-ray powder diffraction, small pieces of crystalline blocks from different parts, along the length of ingot for different compounds, are crushed into respective powder and are recorded in the Bragg-Brentano configuration at room temperature. The Rietveld refinements of the RT powder patterns of respective chalcogenides are illustrated in Fig. 1(a). The calculated patterns (continuous black lines) are matched well with the observed profiles (red solid lines) and absence of any parasitic peaks establishes phase purity of different single crystals. Pseudo-Voigt function and linear interpolation as implemented in FullProf Suite program (2. 05), version: July-2011[25] are used for defining the peak shapes and pattern backgrounds respectively. The growth direction of the mechanically exfoliate shinny crystalline faces having thicknesses of ~135 μm for $Bi_2Te_3$, ~61 μm for $Pb_2Bi_2Te_3$, ~151 μm for $PbBi_2Te_4$ and ~105 μm for $PbBi_4Te_7$ are investigated (for more information about thickness measurement see Fig. S1 in supplementary). The good crystallinity in all the samples is proved by their sharp *c* axes *(00l)* Bragg reflections in their respective x-ray diffraction patterns and are illustrated in Fig. 1(b).

Refined lattice parameters and Wyckoff positions are modelled by VESTA 3.1.8, electronic and structural analysis software package [26] and are illustrated in Fig. 2(a-d). The structural refinement suggests that, the layered rhombohedral structure with space group R-3m (166) for bulk $Bi_2Te_3$, $Pb_2Bi_2Te_3$ and $PbBi_2Te_4$; whereas $PbBi_4Te_7$ belongs to P-3m1 (164). With nearly same lattice parameter *a*= 4.461 Å, the unit cell of bulk $Pb_2Bi_2Te_3$ [$(Pb_2)(Bi_2Te_3)$] contains three bilayers of Pb and three $Bi_2Te_3$ QLs, which are alternately arranged along the crystallographic *c* axis in the sequence: -(Te(2)-Bi-Te(1)-Bi-Te(2))-(Pb-Pb)-; leads to elongation of

$c$ axis from ~30.702 Å for $Bi_2Te_3$ to ~41.947 Å. Covalent interactions between the 6$p$ orbitals of Pb atoms in the Pb bilayers make the slab stable. $PbBi_2Te_4$ [(PbTe)($Bi_2Te_3$)] unit cell consists of three seven-layers packets: -(Te(2)-Bi-Te(1)-Pb-Te(1)-Bi-Te(2))- i.e. the seven monoatomic layers are centro-symmetrical with reference to Pb; Pb acts as inversion center. The insertion of PbTe layers in $Bi_2Te_3$ QLs leads to the elongation of $c$ axis (~42.003 Å). Five atomic layers of $Bi_2Te_3$ block are separated from $PbBi_2Te_4$ block by van-der-Waals gaps in the hexagonal unit cell of $PbBi_4Te_7$, consists of a total twelve atomic layers arranged periodically along the crystallographic $c$ axis that eventually leads to the evolution of three dimensional crystal structure.

In order to investigate the modifications of lattice vibrations and phonon dynamics of $Pb_2Bi_2Te_3$, $PbBi_2Te_4$ and $PbBi_4Te_7$ caused by the insertion of Pb and PbTe layers in $Bi_2Te_3$ in the temperature range from RT to ~93 K, we start with the virgin $Bi_2Te_3$. At room temperature, four phonon modes at 94, 104, 124 and 142 cm$^{-1}$ respectively have been identified and are illustrated in Fig. 3(a). The 104 cm$^{-1}$ i.e., $E_g^2$ mode and 142 cm$^{-1}$ i.e., $A_{1g}^2$ mode are related to the in-plane and the out-of-plane Bi, Te vibrations respectively. The 124 cm$^{-1}$ i.e., $A_{1u}^1$ mode is IR active in bulk and corresponds to the out-of-plane vibrations of the Bi, Te atoms.[17] Teweldebrhan $et$ $al.$[27] have observed 120 cm$^{-1}$ mode and described it as symmetry breaking in atomically thin films of $Bi_2Te_3$. Blue shift is observed for all the modes in the temperature range from RT to 93 K. The evaluation of out-of-plane $A_{1u}^1$ and $A_{1g}^2$ modes relative to the in-plane $E_g^2$ mode with decreasing temperature signify the reduction of strain and reinforcement of the out-of-plane Bi-Te vibrations relative to the in-plane motion. The inset in Fig. 3(a) has illustrated the Lorentzian line shape fitting for the observed modes at 93 K. Fig.

4(a) has illustrated the linear fit of temperature variation of the $A_{1g}^2$ mode having with a slope of 0.01264 ± 0.00083 cm$^{-1}$K$^{-1}$ which is in good agreement with Park *et al.*[28]

Fig. 3(b) has illustrated the temperature dependent Raman spectra for Pb$_2$Bi$_2$Te$_3$. At room temperature we have identified Raman modes at 94, 104 and 124 cm$^{-1}$ respectively along with a notch at 142 cm$^{-1}$. The mode at 124 cm$^{-1}$ is relatively broad; multi-peaks fit by considering Lorentzian line shape shows the existence of two distinct modes at ~124 cm$^{-1}$ i.e., $A_{1u}^1$ and ~132 cm$^{-1}$ i.e., $A_{1u}^2$ respectively. With lowering the temperature these two modes become completely distinct at ~183 K. This observation can be explained by considering thermal noise effect which get suppress with decreasing temperature from RT and finally at low temperature, the phonon scattering effect becomes insignificant, makes the modes prominent and distinct. The notch like 142 cm$^{-1}$ i.e., $A_{1g}^2$ mode evolves to a shape of peak at ~263 K and after that its intensity changes gradually till the end of the temperature variation (i.e., lowest temperature achieved) and are illustrated in the main panel of the Fig. 3(b). At 93 K the modes appear at 94, 106, 126, 137 and 145 cm$^{-1}$ respectively. Theoretically predicted $A_{1u}^2$ mode for Bi$_2$Te$_3$ is identified distinctly at ~137 cm$^{-1}$ for Pb$_2$Bi$_2$Te$_3$ at low temperature. Xu *et al.* [20] have identified a very negligible signature of 132 cm$^{-1}$ mode and explained as weak bonding effects at the boundaries of QLs. In our case, the $A_{1u}^2$ mode at room temperature is very hard to observe but with decreasing temperature this becomes distinct. It can be justified by considering the chemical pressure as induces by Pb layers insertion in Bi$_2$Te$_3$ QLs and results in increasing strength of interactions at the boundaries of Pb layers and Bi$_2$Te$_3$ QL slabs. It is to be noted that, simultaneously the strength of out-of-plane Bi-Te vibrations increases with decreasing the temperature from RT. These two combined effects can be attributed to the evolution of $A_{1u}^2$ mode. For both Bi$_2$Te$_3$ and Pb$_2$Bi$_2$Te$_3$ the $E_g^2$ mode appears at 104 cm$^{-1}$ infers that the nature of covalent interactions and corresponding in-plane motion of Bi-Te are same for both crystals. At RT, the 104 cm$^{-1}$ mode takes the highest value, but as we go down the temperature to ∼93 K, the 124 cm$^{-1}$ peak evolves gradually and becomes the highest intense mode at ~93 K reflects the fact that the out-of-plane Bi-Te vibrations become less restrain over the in-plane Bi-Te vibrations at low temperatures.

Now we focus our attention on the PbBi$_2$Te$_4$ Raman modes at 95, 105, 142 cm$^{-1}$ and two small humps at 124 and 131 cm$^{-1}$ respectively that are observed at room temperature are illustrated clearly in Fig. 3(c). The exact peak positions for these two humps are obtained by deconvoluting the spectra using Lorentzian function. The shapes of the humps gradually evolve at the cost of lowering the temperature and beome distinct at ~183 K. In total, we have identified five modes at low temperature. The intensity of $E_g^2$ mode at 105 cm$^{-1}$ decrease gradually relative to the $A_{1u}^1$ mode at 124 cm$^{-1}$ as we decreases the temperature and $A_{1u}^1$ becomes the highest one at 93 K. The evolution of humps and increasing intensity of $A_{1u}^1$ relative to $E_g^2$ establish the enhancement of out-of-plane Bi-Te vibrations compare to in-plane Bi-Te vibrations at low temperatures. The existance of the 131 cm$^{-1}$ i.e., $A_{1u}^2$ mode may indicate the modification of the van-der-Waals interactions at the boundaries of QLs. The $E_g^2$ mode for PbBi$_2$Te$_4$ arises at 105 cm$^{-1}$ but for Bi$_2$Te$_3$ and Pb$_2$Bi$_2$Te$_3$, the peak position is same (∼104 cm$^{-1}$). It indicates that, the insertion of PbTe atomic layers in $Bi_2Te_3$ slightly modify the covalent interactions between Bi-Te as the Pb atom is at the centro-symmetrical position and is connected with the Te atom and another end of Te is connected with the Bi. Thus Pb in turn modifies the covalent interactions in Bi-Te atoms pair.

Fig. 3(d) has illustrated the temperature variation of Raman modes of $PbBi_4Te_7$. Two distinct modes $E_g^2$ at 104 cm$^{-1}$ and $A_{1u}^1$ at 124 cm$^{-1}$ are identified at RT. With decreasing temperature the $A_{1u}^1$ mode evolves gradually and its intensity becomes almost equal to $E_g^2$ at ~163 K. A new mode is observed at ~147 cm$^{-1}$ at ~113 K and becomes distinct at the lowest measure temperature here, which can be explained in terms of the enhancement of the out-of-plane vibrations over in-plane Bi-Te vibrations at lower temperatures. The maturity of $A_{1g}^2$ mode at 142 cm$^{-1}$ is accompanied by the enhancement of the $A_{1u}^1$ mode which becomes the most intense Raman mode at ~113 K. Due to poor signal to noise ratio the spectra observed in case of $PbBi_4Te_7$, the mode at ~95 cm$^{-1}$ is not well resolved and the appearance of new mode at 147 cm$^{-1}$ due to insertion of PbTe layers is not clearly visible, therefore only three modes at 111, ~130 and 147 cm$^{-1}$ are discussed here.

The observed ~95 cm$^{-1}$ mode for $Bi_2Te_3$, $Pb_2Bi_2Te_3$ and $PbBi_2Te_4$ more or less behaves in the same way. That means, with lowering the temperature; they remain un-shifted but their shapes evolve gradually and are broadened as clearly observed in the Fig. 3(a-c). This is in good agreement with the Wang et al.[14], who have performed their experiment on MBE grown thin film and Raman measurement under ultra-high vacuum, explains 95 cm$^{-1}$ mode as the surface phonon mode. But the origin of this mode is not beyond debate, He et al.[29] have argued that, the inversion symmetry breaking for their nano-plates whereas, Jian-Hua et al.[30]

have described it as surface oxidation effect. Very recently Fernández et al.[31] reports that, Te clusters in Te rich $Bi_2Te_3$ leads to 88 cm$^{-1}$ mode. We have found 95 cm$^{-1}$ mode in theoretical studies in DFT calculation.

The peak position and full width half maxima (FWHM) as obtain from Lorentzian line shape fitting of individual modes as a function of temperature for each specimen are illustrated in Fig. 4(a-d). Temperature dependency of peak positions for the observed $Bi_2Te_3$ and $Pb_2Bi_2Te_3$ Raman modes are fitted well with the consideration of anharmonic phonon-phonon coupling along with thermal expansion.[28] It provides a reasonably precise depiction of the phonon temperature dependence in some semiconductors and diamond.[32, 33] Let, $\omega_0$ is the bare harmonic frequency, $\chi^{/}T$ is due to the lattice thermal expansion, and $\chi^{//}T^2$ is the anharmonic phonon-phonon coupling term. Therefore, the temperature dependence of the phonon frequency can be given by:

$$\omega = \omega_0 + \chi^{/}T + \chi^{//}T^2,$$

where $\chi^{/}$ and $\chi^{//}$ are first order and second order temperature coefficients respectively. For $PbBi_2Te_4$ and $PbBi_4Te_7$ with decreasing the temperature from RT, the modes shift towards the higher wave number side. It infers that, the strength of covalent interactions between Bi-Te increases and hence force constant of the interacting forces increase which in turn produces blue shift in Raman modes. As a result of insertion of Pb and PbTe layers in $Bi_2Te_3$ the out-of-plane Bi-Te vibrations become less restrained compare to the in-plane vibrations as temperature decreases from RT to 93 K. Increase of slope of the $A_{1g}^2$ mode is tabulated in Table 1.

## Density functional theory simulations

We have also performed First principles simulations and computed the Raman modes and frequencies for all the samples. All the calculations have been performed within the framework of density functional theory using the Quantum espresso 5.3.0 code.[34] We have used the tested PBEsol pseudo potential for Bi, Pb and Te from THOES library.[35] The plain waves and charge density cut-off is set to 55 Ry and 500 Ry respectively and a Monkhorst-Pack [36] grid of *k*-points is chosen according to the symmetry of the crystal. The phonon properties are calculated using the phonopy package [37] which is used for post-processing the data from Quantum espresso. A super-cell and finite difference (FD) approach are used to calculate the dynamical matrix. Fig. 5(a-c) represents the density of states (DOS) plots for $Pb_2Bi_2Te_3$, $PbBi_2Te_4$ and $PbBi_4Te_7$ respectively. The electronic properties as display in Fig. 5(a-c) reveal that $PbBi_2Te_4$ and $PbBi_4Te_7$ exhibit insulating character whereas $Pb_2Bi_2Te_3$ possesses semi-metallic signature. $Pb_2Bi_2Te_3$ is analogous to $Bi_4Te_3$, which is reported to be semi metallic.[21] So it is expected that $Pb_2Bi_2Te_3$ may behave as semi-metallic. We can infer that due to the insertion of Pb slab within the $Bi_2Te_3$ slabs the narrow gap (~170 meV) of $Bi_2Te_3$ vanishes and system becomes metallic, this finding is consistent with literature report.[22] The experimentally observed Raman modes and the frequencies for the bulk structures of $PbBi_2Te_4$, $PbBi_4Te_7$ are the same as that obtained from bulk $Bi_2Te_3$. The main reason is the higher atomic weight of the Pb atom which reduces the mode frequency under harmonic approximations (since, $\omega = \sqrt{k/M}$), which is clearly seen from the PPDOS and are illustrated in Fig. 6(a-c). The phonon modes from Pb only contribute to the lower side of the spectrum and not of those which are a Raman signature. Theoretically calculated mode frequencies along with the experimentally observed frequencies are tabulated in Table 2. For bulk $PbBi_2Te_4$,

Raman modes match nicely with the experimental observations except slight deviation for low frequency modes. The low frequency mode at ~95 cm$^{-1}$ obtained in experiment comes at little lower frequency (81 cm$^{-1}$). This deviation between experiment and theory is very much reasonable as Cheng *et al.*[18] have got around 8-10 cm$^{-1}$ difference between experiment and theory even after including spin-orbit interactions. For Bi$_2$Te$_3$ film, the frequencies of most of the modes match nicely with the experimental observations except for the highest mode which is overestimated and can be lowered with the incorporation of spin-orbit interactions.[18] The band structure of Bi$_2$Te$_3$ film is resolved correctly when including spin-orbit effect and the Dirac cone is observed at the Γ point in the Brilliuon zone and is illustrated in Fig. S3 in supplementary. The errors in mode frequencies in DFT are due to the sensitivity of calculation with the external stresses which is seen in previous calculations.

## Conclusions

A comprehensive study of temperature dependent Raman spectra in layered single crystals of Pb$_2$Bi$_2$Te$_3$, PbBi$_2$Te$_4$ and PbBi$_4$Te$_7$ has been presented. The general blue shift trends with decrease in temperature are observed for all studied materials due to their similar lattice structures. The overall agreement between theoretically calculated DFT and experimentally observed modes at 93 K is quite good for all the samples. In the phonon density of states, Pb atoms contribute mainly in the lower frequency region whereas higher frequency contribution comes from Te and Bi. We found that at temperatures above 150 K, the frequency of the $A_{1g}^2$ phonon change linearly with temperature. Whereas below 150 K the evaluation of $A_{1g}^2$ modes are best fitted by nonlinear polynomial and are understood by the inclusion of anharmonic phonon-phonon coupling for $Pb_2Bi_2Te_3$. Analogues electronic, crystal structure and Raman spectra of $Pb_2Bi_2Te_3$ and $Bi_4Te_3$ indicate that $Pb_2Bi_2Te_3$ may be a promising candidate of topological semimetal, though the ARPES measurement is necessary to confirm the topological property in $Pb_2Bi_2Te_3$. The experimentally observed Raman modes for $PbBi_2Te_4$ and $PbBi_4Te_7$ at the same frequencies as that of $Bi_2Te_3$ explain theoretically under the harmonic approximations by lattice thermal expansion and enhancement of interaction stiffness affects. The insertion of Pb and PbTe layers in $Bi_2Te_3$ introduces chemical pressure in the unit cell of $Bi_2Te_3$ and modifies the interactions at the QLs boundary results in observation of $A_{1u}^2$ mode. Furthermore at the lowest measured temperature, the $A_{1u}^1$ mode becomes the highest one reflects the enhancement of the out-of-plane Bi-Te vibrations relative to the in-plane Bi-Te vibrations. Similar observation is made for structural evolution of $Bi_2Te_3$ to $PbBi_4Te_7$.

## ACKNOWLEDGEMENTS

Pradip Das and Priyanath Mal acknowledge the Department of Science and Technology for financial support through project no. SR/FTP/PS-197/2012. Authors of GGV also thank to UGC, Govt. of India for supporting the Department of Pure and Applied Physics through UGC SAP DRS –I and FIST Level – I programs respectively. Brahmananda Chakraborty would like to thank Dr. N. K. Sahoo for support and encouragement. Brahmananda Chakraborty would also like to thank the staff of BARC computer division for supercomputing facility.

†pradipd.iitb@gmail.com

*brahma@barc.gov.in

# References

1  M. Z. Hasan and C. L. Kane, Rev. Mod. Phys. **82**, 3045-3067 (2010).

2  X. -L. Qi and S. -C. Zhang, Rev. Mod. Phys. **83**, 1057-1110 (2011).

3  J. E. Moore, Nature (London) **464**, 194 (2010).

4  M. S. König, C. Wiedmann, A. Brüne, H. Roth and L. Buhmann, Science **318**, 766 (2007).

5  D. Pesin and A. H. MacDonald, Nature Mater. **11**, 409 (2012).

6  D. A. Wright, Nature **181**, 834 (1958).

7  M. Saleemi, M. S. Toprak, S. Li, M. Johnssonb and M. Muhammed, J. Mater. Chem. **22**, 725 (2012).

8  L. Zhang and D. J. Singh, Phys. Rev. B **81**, 245119 (2010).

9  T. V. Quang and M. Kim, Journal of the Korean Physical Society **68**, 393 (2016).

10  J. J. Zhou, W. Feng, Y. Zhang, S. A. Yang and Y. Yao, Sci. Rep. **4**, 3841 (2014).

11  S. Barua, K. P. Rajeev and A. K. Gupta, J. Phys.: Condens. Matter **27**, 015601 (2015).

12  K. Kuroda, H. Miyahara, M. Ye, S. V. Eremeev, Yu. M. Koroteev, E. E. Krasovskii, E. V. Chulkov, S. Hiramoto, C. Moriyoshi, Y. Kuroiwa, K. Miyamoto, T. Okuda, M. Arita, K. Shimada, H. Namatame, M. Taniguchi, Y. Ueda and A. Kimura, Phys. Rev. Lett. **108**, 206803 (2012).

13  T. Okuda, T. Maegawa, M. Ye, K. Shirai, T. Warashina, K. Miyamoto, K. Kuroda, M. Arita, Z. S. Aliev, I. R. Amiraslanov, M. B. Babanly, E. V. Chulkov, S. V. Eremeev, A. Kimura, H. Namatame and M. Taniguchi, Phys. Rev. Lett. **111**, 206803 (2013).

14      C. Wang, X. Zhu, L. Nilsson, J. Wen, G. Wang, X. Shan, Q. Zhang, S. Zhang, J. Jia and Q. Xue, Nano Res. **6**, 688 (2013).

15      V. Chis, I. Yu. Sklyadneva, K. A.Kokh, V. A. Volodin and O. E. Tereshchenko and E. V. Chulkov, Phys. Rev. B **86**, 174304 (2012).

16      K. M. F. Shahil, M. Z. Hossain, V. Goyal and A. A. Balandin, J. Appl. Phys. **111**, 054305 (2012).

17      J. Yuan, M. Zhao, W. Yu, Y. Lu, C. Chen, M. Xu, S. Li, K. P. Loh and Q. Bao, Materials **8**, 5007 (2015).

18      W. Cheng and S. F. Ren, Phys. Rev. B **83**, 094301 (2011).

19      T. Hirahara, G. Bihlmayer, Y. Sakamoto, M. Yamada, H. Miyazaki, S. Kimura, S. Blugel and S. Hasegawa, Phys. Rev. Lett. **107**, 166801 (2011).

20      H. Xu, Y. Song, W. Pan, Q. Chen, X. Wu, P. Lu, Q. Gong and S. Wang, AIP Advances **5**, 087103 (2015).

21      J. W. G. Bos, H. W. Zandbergen, M. -H. Lee, N. P. Ong and R. J. Cava, Phys. Rev. B **75**, 195203 (2007).

22      H. Lind, S. Lidin and U. Häussermann, Phys. Rev. B **72**, 184101 (2005).

23      T. Valla, H. Ji, L. M. Schoop, A. P. Weber, Z. -H. Pan, J. T. Sadowski, E. Vescovo, A. V. Fedorov, A. N. Caruso, Q. D. Gibson, L. Müchler, C. Felser and R. J. Cava, Phys. Rev. B **86**, 241101(R) (2012).

24      V. Russo, A. Bailini, M. Zamboni, M. Passoni, C. Conti, C. S. Casari, A. L. Bassi and C. E. Bottani, J. Raman Spectrosc. **39**, 205 (2008).

25      J. Rodríguez-Carvajal, An introduction to the program France: FullProf 2000; **2001**

26      K. Momma and F. Izumi, J. Appl. Crystallogr. **44**, 1272 (2011).

27      D.Teweldebrhan, V. Goyal and A. A. Balandin, Nano Lett. **10**, 1209 (2010).

28  D. Park, S. Park, K. Jeong, H. –S. Jeong, J. Y. Song and M. –H. Cho, Sci. Rep. **6**, 19132 (2016).

29  R. He, Z. Wang, R. L. J. Qiu, C. Delaney, B. Beck, T. E. Kidd, C. C. Chancey and X. P. A. Gao, Nanotechnology **23**, 455703 (2012).

30  G. J. -Hua, Q. Feng, Z. Yun, D. H. -Yong, H. G. -Jin, L. X. -Nan, Y. G. -Lin and D. Ning, Chin. Phys. Lett. **30**, 106801 (2013).

31  C. R. -Fernández, C. V. Manzano, A. H. Romero, J. Martín, M. M. -González, M. M. de Lima Jr and A. Cantarero, Nanotechnology **27**, 075706 (2016).

32  Y. Kim, X. Chen, Z. Wang, J. Shi, I. Miotkowski, Y. P. Chen, P. A. Sharma, A. L. L. Sharma, M. A. Hekmaty, Z. Jiang and D. Smirnov, Appl. Phys. Lett. **100**, 071907 (2012).

33  M. S. Liu, L. A. Bursill, S. Prawer and R. Beserman, Phys. Rev. B **61**, 3391 (2000).

34  P. Giannozzi, S. Baroni, N. Bonini, M. Calandra, R. Car, C. Cavazzoni, D. Ceresoli, G. L. Chiarotti, M. Cococcioni, I. Dabo, A. D. Corso, S. Fabris, G. Fratesi, S. de Gironcoli, R. Gebauer, U. Gerstmann, C. Gougoussis, A. Kokalj, M. Lazzeri, L. M. -Samos, N. Marzari, F. Mauri, R. Mazzarello, S. Paolini, A. Pasquarello, L. Paulatto, C. Sbraccia, S. Scandolo, G. Sclauzero, A. P. Seitsonen, A. Smogunov, P. Umari, R. M. Wentzcovitch, J. Phys.: Condens. Matter. **21**, 395502 (2009).

35  http://theossrv1.epfl.ch/Main/Pseudopotentials

36  H. J. Monkhorst and J. D. Pack, Phys. Rev. B **13**, 5188 (1976).

37  A. Togo and I. Tanaka, Scr. Mater. **108**, 1 (2015).

**Figure captions**

**Figure 1:** (a) Red, black and blue solid lines are observed, calculated and difference pattern profiles respectively, obtained from the refinement of the room temperature powder diffraction data. The green vertical parallel ticks above the difference profile correspond to the Bragg peak position. (b) X-ray diffraction of the mechanically exfoliated single crystalline planes. All the peaks observed in the recorded pattern are indexed according to the growth direction. The single crystalline plane of $Pb_2Bi_2Te_3$ and optical image of the crystalline plane for $PbBi_2Te_4$ are illustrated in the inset (second and third row).

**Figure 2:** Schematic representation of unit cells of (a) $Bi_2Te_3$, (b) $Pb_2Bi_2Te_3$, (c) $PbBi_2Te_4$ and (d) $PbBi_4Te_7$ respectively, looking along crystallographic *a* axis: the solid line represents the unit cell. The bond lengths are in Å.

**Figure 3:** Temperature dependent Raman spectra of (a) 135 μm $Bi_2Te_3$, (b) 61 μm $Pb_2Bi_2Te_3$, (c) 151μm $PbBi_2Te_4$ and (d) 104μm $PbBi_4Te_7$ are recorded with 532 nm laser diode source with a spectral resolution of 1 $cm^{-1}$. The Lorentzian line shape fitting of the RT and 93 K modes are illustrated in the main panel and inset respectively. Blue lines correspond to individual peak fit and that of red lines represents the peak sum.

**Figure 4:** Mode shift and full width half maxima (FWHM) variations with temperature for (a) $Bi_2Te_3$, (b) $Pb_2Bi_2Te_3$, (c) $PbBi_2Te_4$ and (d) $PbBi_4Te_7$. Black, red and green colored axes correspond to Temperature, FWHM and Raman shift respectively. Red open circles and green open square represents FWHM and peak position respectively. Dashed black and blue lines represent the linear fit and quadratic fit of anharmonic phonon-phonon coupling respectively.

**Figure 5:** Total density of states of (a) $Pb_2Bi_2Te_3$, (b) $PbBi_2Te_4$ and (c) $PbBi_4Te_7$ respectively. The dashed black line corresponds to the Fermi level.

**Figure 6:** Partial phonon density of states of (a) $Pb_2Bi_2Te_3$, (b) $PbBi_2Te_4$ and (c) $PbBi_4Te_7$ respectively.

**Table captions**

**Table 1:** Comparison of first order temperature coefficient ($\chi^{/}$) of different Raman modes of $Bi_2Te_3$, $Pb_2Bi_2Te_3$, $PbBi_2Te_4$ and $PbBi_4Te_7$.

**Table 2:** The characteristics Raman frequencies of different specimens calculated theoretically and experimentally observed at 93 K.

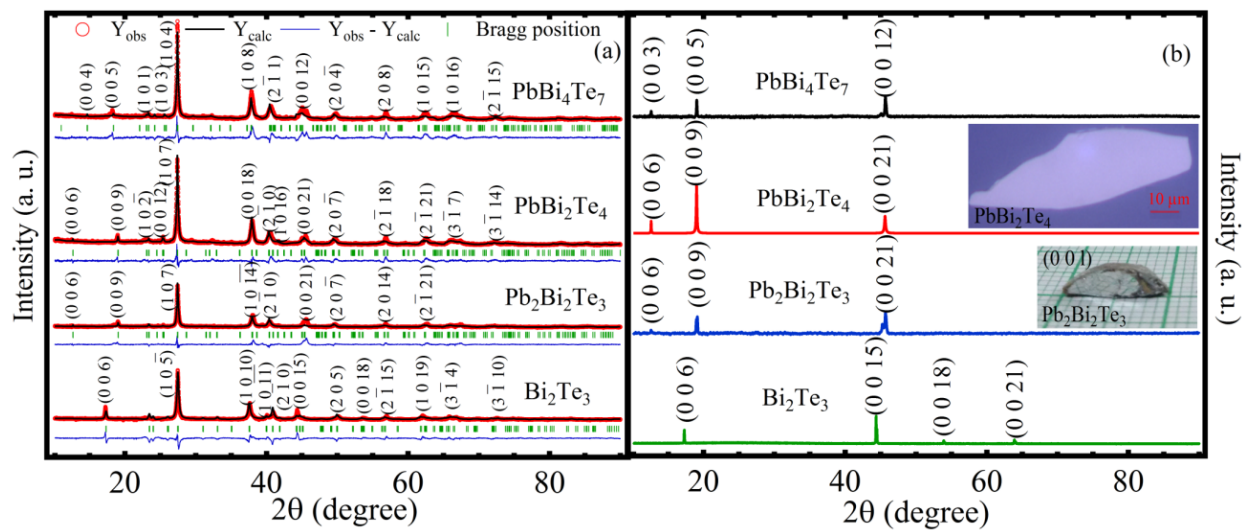

Figure 1

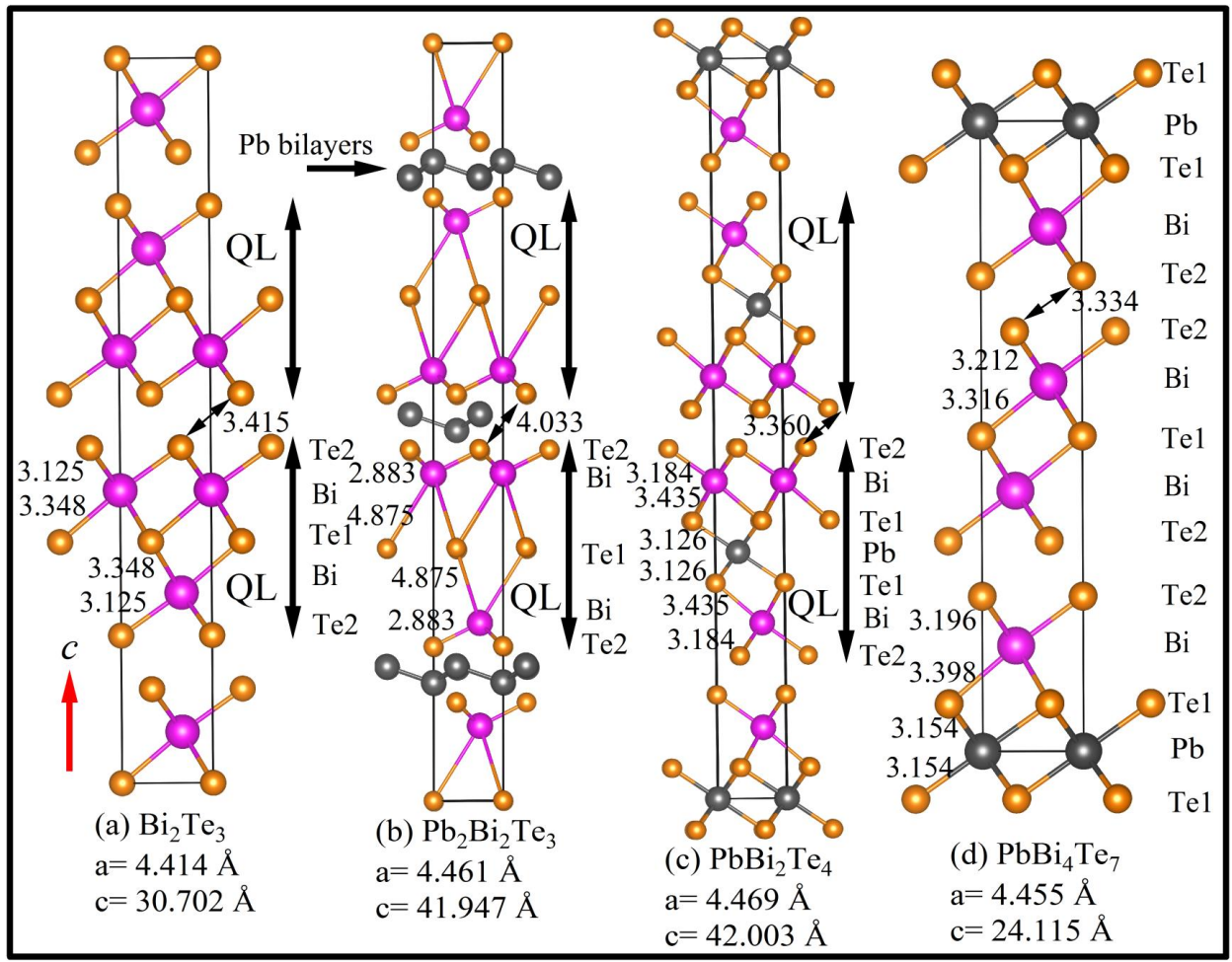

Figure 2

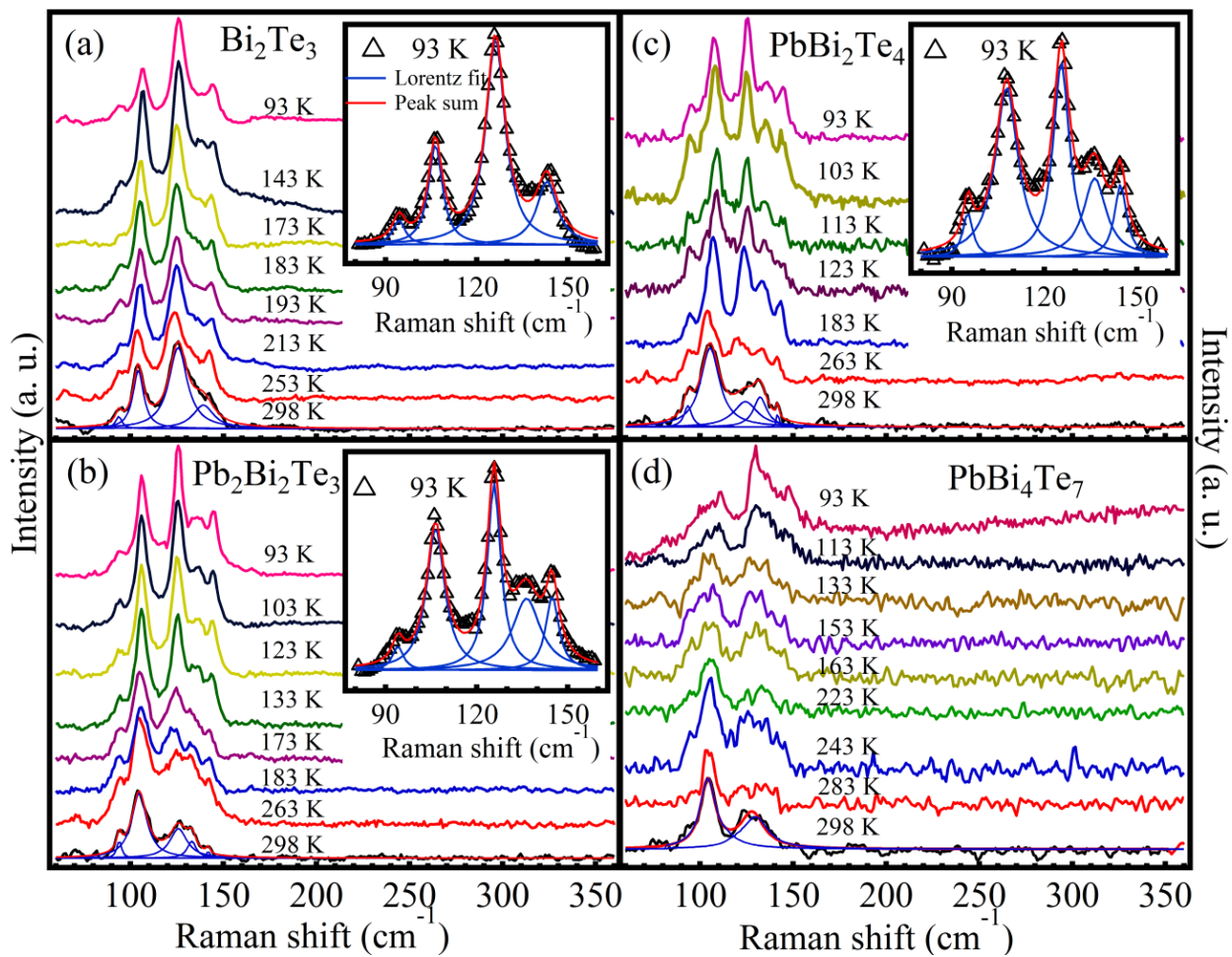

Figure 3

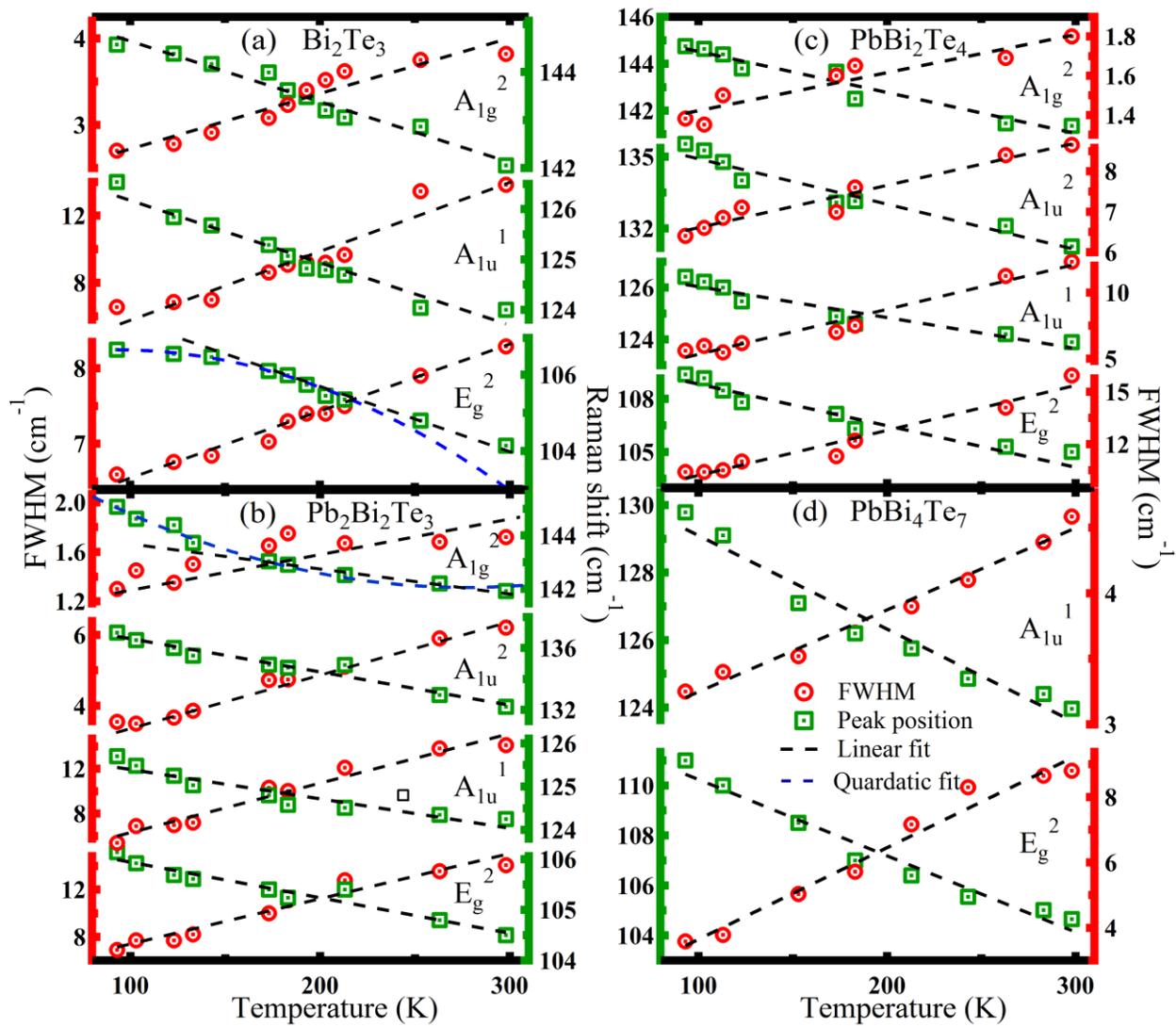

Figure 4

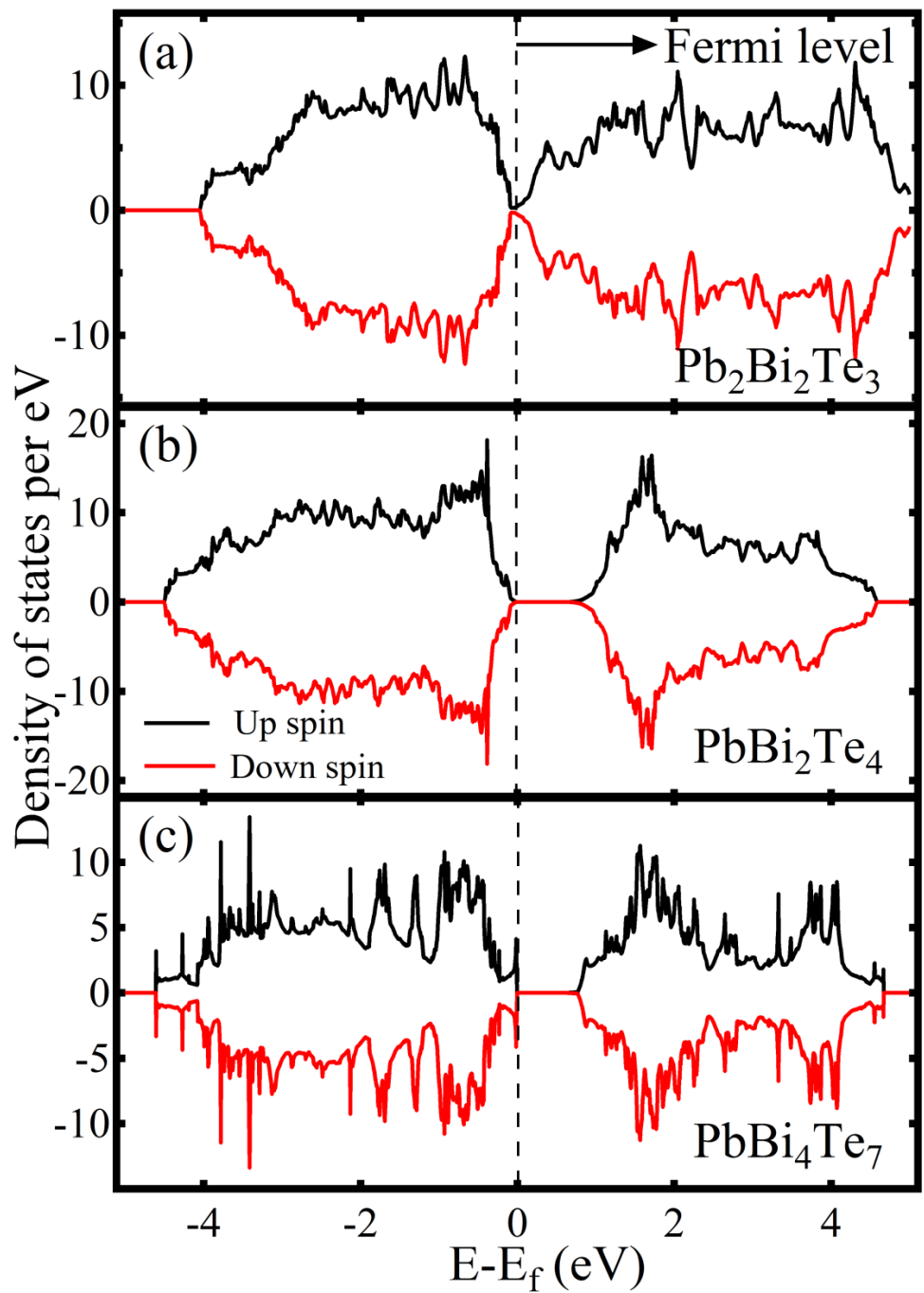

Figure 5

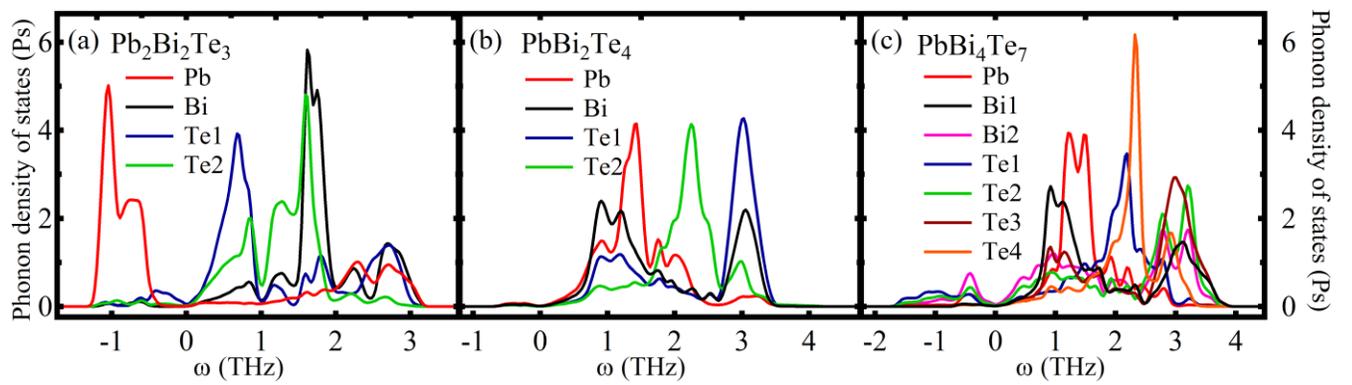

Figure 6

Table 1

| Sample name | Raman modes | $\chi'$: slope of linear fit (cm$^{-1}$K$^{-1}$) |
|---|---|---|
| Bi$_2$Te$_3$ | $E_g^2$ | -0.01317 ± 0.00094 |
| | $A_{1u}^1$ | -0.01252 ± 0.00119 |
| | $A_{1g}^2$ | -0.01264 ± 0.00083 |
| Pb$_2$Bi$_2$Te$_3$ | $E_g^2$ | -0.00704 ± 0.00065 |
| | $A_{1u}^1$ | -0.00678 ± 0.00089 |
| | $A_{1u}^2$ | -0.00568 ± 0.00029 |
| | $A_{1g}^2$ | -0.00869 ± 0.00072 |
| PbBi$_2$Te$_4$ | $E_g^2$ | -0.02331 ± 0.00192 |
| | $A_{1u}^1$ | -0.01195 ± 0.00153 |
| | $A_{1u}^2$ | -0.01748 ± 0.00177 |
| | $A_{1g}^2$ | -0.01933 ± 0.00196 |
| PbBi$_4$Te$_7$ | $E_g^2$ | -0.03051 ± 0.00244 |
| | $A_{1u}^2$ | -0.02771 ± 0.00242 |

Table 2

| Bi$_2$Te$_3$ Mode frequency (cm$^{-1}$) | | Pb$_2$Bi$_2$Te$_3$ Mode frequency (cm$^{-1}$) | | PbBi$_2$Te$_4$ Mode frequency (cm$^{-1}$) | | PbBi$_4$Te$_7$ Mode frequency (cm$^{-1}$) | |
|---|---|---|---|---|---|---|---|
| Theo. | Expt. | Theo. | Expt. | Theo. | Expt. | Theo. | Expt. |
| 27 | ----- | ----- | ----- | ----- | ----- | ----- | ----- |
| ----- | 94 | 94 | 94 | 81 | 95 | 91 | ----- |
| 101 | ~107 | 102 | 106 | 105 | 107 | 105 | 111 |
| 123 | ~127 | ----- | ~126 | 111 | ~127 | 122 | ~130 |
| ----- | ----- | ----- | 137 | 114 | ~136 | 128 | ----- |
| 158 | ~145 | ----- | 145 | ----- | ~145 | ----- | 147 |

# Supplementary information

**Figure S1:** Thickness of (a) Bi2Te3, (b) Pb2Bi2Te3, (c) PbBi2Te4 and (d) PbBi4Te7 single crystal flakes measured through optical microscope.

**Figure S2:** EDAX spectra for (a) $Bi_2Te_3$, (b) $Pb_2Bi_2Te_3$, (c) $PbBi_2Te_4$ and (d) $PbBi_4Te_7$ single crystal flakes.

**Figure S3:** Band structure of $Bi_2Te_3$.

**Figure S4:** Phonon dispersion of $Bi_2Te_3$.

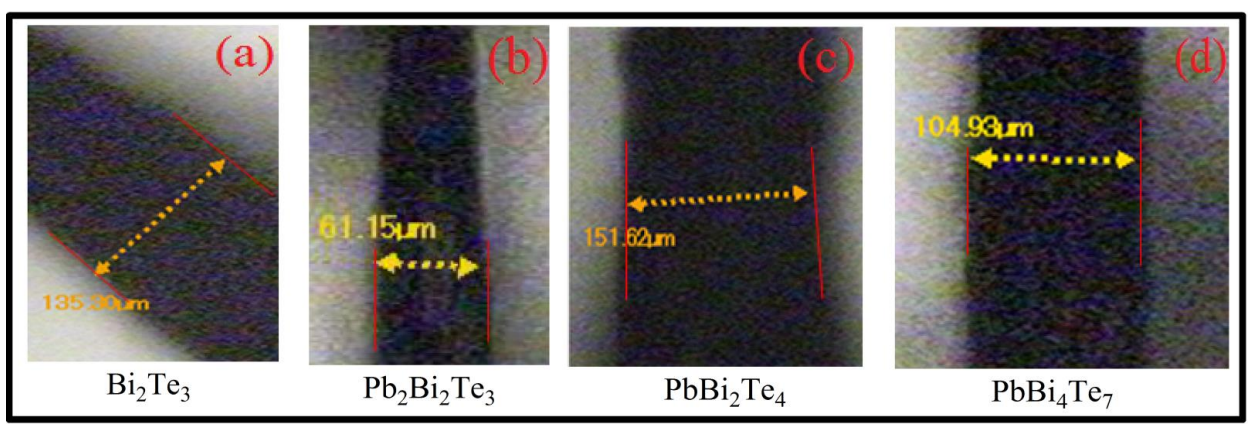

Figure S1

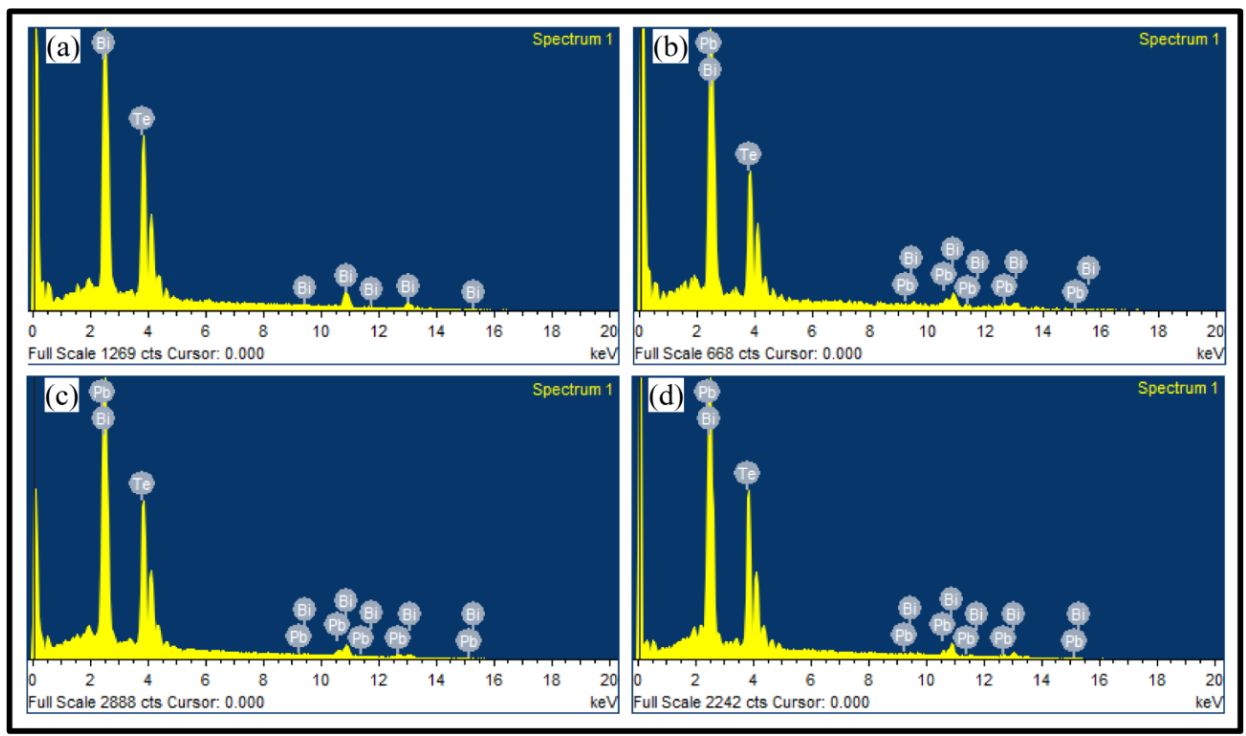

| Specimen | Weight (%) | |
|---|---|---|
| | Theoretical | Experimental |
| Bi$_2$Te$_3$ | Bi: 52.19 | Bi: 52.24 |
| | Te: 47.80 | Te: 47.47 |
| Pb$_2$Bi$_2$Te$_3$ | Pb: 34.0 | Pb: 32.83 |
| | Bi: 34.40 | Bi: 33.43 |
| | Te: 31.50 | Te: 33.74 |
| PbBi$_2$Te$_4$ | Pb: 19.31 | Pb: 19.31 |
| | Bi: 36.57 | Bi: 35.73 |
| | Te: 45.11 | Te: 44.96 |
| PbBi$_4$Te$_7$ | Pb: 10.70 | Pb: 10.10 |
| | Bi: 43.17 | Bi: 43.16 |
| | Te: 46.12 | Te: 46.73 |

Figure S2

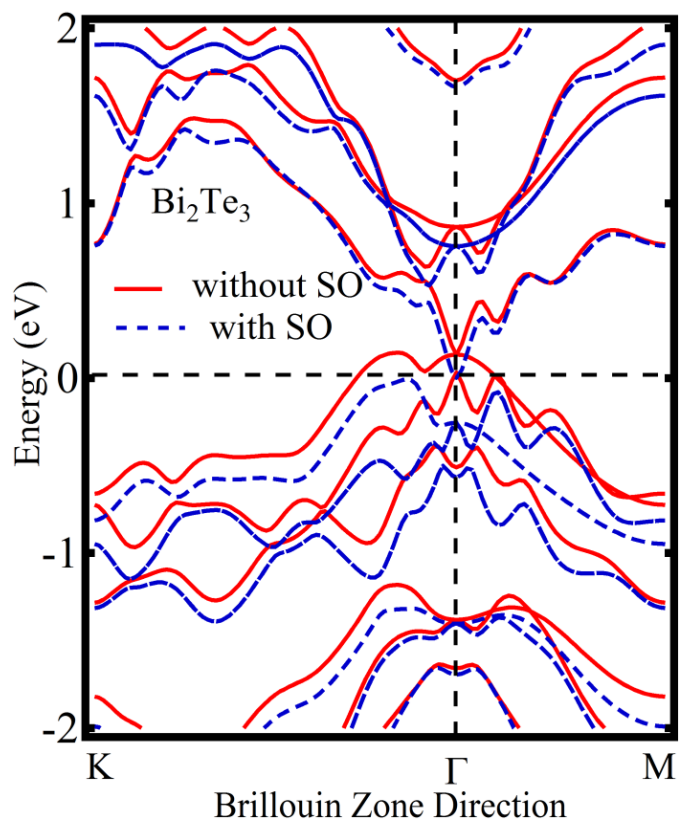

Figure S3

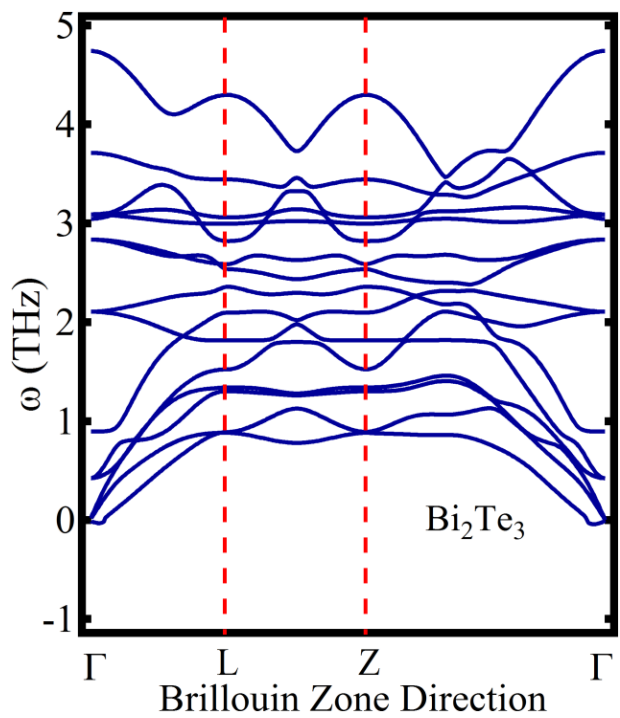

Figure S4